\documentclass[12pt]{article}
\usepackage{amscd}
\usepackage{verbatim}
\usepackage{amssymb}

\begin{document}
\vskip 0 true cm
\flushbottom
\begin{center}
\vspace{24pt}
{ \large \bf Irreversibility of World-sheet Renormalization Group Flow} \\
\vspace{30pt}
{\bf T Oliynyk}$^{\dag}$ \footnote{oliynyk@math.ualberta.ca}, {\bf V Suneeta}$^{\ddag}$ \footnote{svardara@phys.ualberta.ca},
{\bf E Woolgar}$^{\dag,\ddag}$ \footnote{ewoolgar@math.ualberta.ca}

\vspace{24pt}
{\footnotesize

$^\dag$ Dept of Mathematical Sciences, University of Alberta,\\
Edmonton, AB, Canada T6G 2G1.\\
$^\ddag$ Theoretical Physics Institute, University of Alberta,\\
Edmonton, AB, Canada T6G 2J1.}
\end{center}
\date{\today}
\vspace{48pt}
\begin{center}
{\bf Abstract}
\end{center}
\noindent We demonstrate the irreversibility of a wide class of world-sheet
renormalization group (RG) flows to first order in $\alpha'$ in string theory. Our
techniques draw on the mathematics of Ricci flows, adapted to asymptotically flat
target manifolds. In the case of somewhere-negative scalar curvature (of the target
space), we give a proof by constructing an entropy that increases monotonically along
the flow, based on Perelman's Ricci flow entropy. One consequence is the absence of
periodic solutions, and we are able to give a second, direct proof of this. If the
scalar curvature is everywhere positive, we instead construct a regularized volume to
provide an entropy for the flow. Our results are, in a sense, the analogue of
Zamolodchikov's $c$-theorem for world-sheet RG flows on noncompact spacetimes (though
our entropy is not the Zamolodchikov $C$-function).


\newpage
\noindent For a wide class of 2-dimensional quantum field
theories, Zamolodchikov's $c$-theorem \cite{Z} demonstrates the
irreversibility of renormalization group (RG) flow. In particular,
there exists a function defined on the space of 2D renormalizable
field theories, the $C$-function, which decreases along RG
trajectories
and is stationary only at RG fixed points.
The fixed points of the RG flow are conformal field theories. The
$C$-function equals the central charge of the conformal field
theory at the fixed points. However, as shown by Polchinski
\cite{Pol}, the $c$-theorem is not valid for a very important
class of 2-dimensional quantum field theories of relevance to
string theory: the world-sheet nonlinear sigma model on {\em
noncompact} target spaces.
There exists no general proof that RG
flows of world-sheet sigma models are irreversible. Indeed,
violations of irreversibility
are known for other kinds of field theory RG flows (cf.\
\cite{niemi} and references therein). A recent focus of study in
this area has been RG flows of sigma models with target spaces
that are 2-dimensional and noncompact, and the question of whether
the {\em mass} of the target spacetime changes monotonically along
the flow has been investigated \cite{APS,GHMS,HMT,Bakas}. However,
the mass at infinity does not change {\em along} the flow; it
changes
only at the final fixed point.

In this paper, we demonstrate the irreversibility of world-sheet RG flow to first order
in $\alpha'$ on complete, asymptotically flat Riemannian manifolds (or static slices of
spacetimes) when all fields other than the metric and the dilaton are set to zero. In
addition to proving that this class of flows does not contain periodic solutions, we
construct an {\em entropy} (by which we mean a Liapunov function) that increases
monotonically along the flow. Thus, in a sense, our result generalises Zamolodchikov's
$c$-theorem to a wide class of string theory world-sheet RG flows on noncompact
spacetimes. Our entropy is not, however, the $C$-function. The $C$-function for the
world-sheet sigma model was computed by Tseytlin \cite{Tseytlin}, who showed that it
could be obtained as a generalized transform of the low-energy string effective action
divided by manifold volume. Thus for the class of RG flows on noncompact spacetimes
that we are interested in, the $C$-function is zero and cannot serve as an entropy
along the flow.

We caution that we work only to first order in $\alpha'$, the square of the string
scale. As is well-known for compact target spaces with $R>0$, for example, higher order
stringy corrections cannot be neglected. For flows where even just the second order
corrections in $\alpha'$ to the beta functions become significant, we are presently
unable to rule out periodic solutions using our methods. This is because our techniques
draw on the theory of quasi-linear differential equations, which does not apply (at
least not straightforwardly) to higher-order RG flow.

To first order in $\alpha'$, the RG equations for the (string frame) metric $g_{ab}$
and the dilaton $\Phi_{D} = \Psi/2$ are
\begin{eqnarray}
\frac{\partial}{\partial \tau}g_{ij} &=& - \alpha^{'}
(R_{ij} + \nabla_i \nabla_j \Psi) \ ,\label{eq1} \\
\frac{\partial}{\partial \tau}\Psi &=& \frac{\alpha^{'}}{2} (\Delta \Psi - \vert \nabla
\Psi \vert^2)\ . \label{eq2}
\end{eqnarray}
Here $\tau$ is the logarithm of the world-sheet RG scale. The RG equation for the
metric first appeared in the 1970s \cite{EH,Friedan}. Subsequent papers \cite{FT}
generalised this analysis to include the effects of other background fields like the
dilaton and the antisymmetric tensor field.

The form of these equations is not diffeomorphism invariant. A
$\tau$-dependent diffeomorphism generated by $\xi_i = \frac{\alpha^{'}}{2} \nabla_i
\Psi$ decouples the metric flow from the dilaton field, so (\ref{eq1}) becomes
\begin{equation}
\frac{\partial}{\partial \tau}g_{ij} = - \alpha^{'} R_{ij} \ , \label{eq3}
\end{equation}
This equation, which we call the {\it Hamilton gauge} RG equation for the metric,
arises in mathematics as a tool in Hamilton's programme \cite{Hamilton} to address
Thurston's geometrization conjecture for closed 3-manifolds, where it is called the
{\it Ricci flow}.

Herein we consider instead asymptotically flat manifolds with
asymptotic structure fixed along the flow. We take such metrics to satisfy
\begin{eqnarray}
(g_{ij}(\tau)-\delta_{ij}) &\in& \mathcal{O} (1/r^{D-2-\epsilon})\ , \nonumber\\
\partial_k g_{ij}(\tau) &\in& \mathcal{O} (1/r^{D-1-\epsilon})
\ , \label{eq4}\\
\partial_k\partial_l g_{ij}(\tau) &\in& \mathcal{O} (1/r^{D-\epsilon})
\ , \nonumber
\end{eqnarray}
and so on up to at least 4 derivatives, uniformly in $\tau$ for any $\epsilon>0$.
(Local existence of such solutions will be presented elsewhere.)

In the sequel, we will study the irreversibility of (\ref{eq3}). This implies the
irreversibility of the system (\ref{eq1},\ref{eq2}), but is more general. To see this,
observe that the system (\ref{eq1},\ref{eq2}) is the pullback along $\xi_i$ of the
system comprised of (\ref{eq3}) and the Hamilton gauge dilaton equation
\begin{equation}
\frac{\partial}{\partial \tau}\Psi = \frac{\alpha^{'}}{2} \Delta
\Psi \ .\label{eq5}
\end{equation}
By the maximum principle for parabolic equations \cite{DEbook}, (\ref{eq5}) has only
monotonic solutions (if, say, $\Psi\to const$ at spatial infinity). In consequence, the
flows (\ref{eq3},\ref{eq5}) and, therefore, (\ref{eq1},\ref{eq2}) are irreversible
unless the dilaton is constant: if it is, then the question reduces to whether
(\ref{eq3}) has reversible flows. But this question is interesting even when the
dilaton is not constant, and this more general case entails no addition burden since
(\ref{eq3}) is independent of the dilaton.

Flows that are periodic up to diffeomorphism are called {\em breathers}. More
precisely, a solution of (\ref{eq3}) is a {\em steady breather} if, for some $\tau_1 <
\tau_2$ and a diffeomorphism $\phi$,
\begin{equation}
g(\tau_1) = \phi^{*}g(\tau_2) \ . \label{eq6}
\end{equation}
{\em Expanding} and {\em shrinking} breathers are defined by including as well an
overall rescaling in (\ref{eq6}), but cannot occur for flows through asymptotically
flat spaces with Euclidean structure at infinity fixed in $\tau$, so we do not consider
them further (though they can produce interesting physics; cf \cite{GHMS}). We will
demonstrate that asymptotically flat breathers occur only in the trivial case in which
the metrics {\em all} along the flow are related by diffeomorphisms and the breather is
a {\em steady Ricci soliton}.

Let $R$ be the scalar curvature of the target manifold. From
(\ref{eq3}) we can derive its flow, which is given by
\begin{equation}
\frac{\partial R}{\partial \tau}=\frac{\alpha'}{2}\left (\Delta
 R +2R_{ij}R^{ij} \right )\ .
\label{eq7}
\end{equation}
Our problem separates into cases according to whether $R$ can be somewhere negative
along the flow, never negative but somewhere zero, or always positive. In Sections 1
and 2, we will adapt Perelman's recently discovered Ricci flow entropy for compact
manifolds \cite{Perelman}) to the asymptotically flat case and use this entropy to rule
out breathers if $R<0$ somewhere. This entropy is not useful if $R\ge 0$ everywhere
along the flow, so in Section 3 we give an entirely different entropy-type argument
based on regularized volume. In an appendix, we give alternative, direct (entirely
non-entropic) geometric arguments based on work of Ivey \cite{Ivey} to rule out
nontrivial breathers if $R\le 0$ somewhere.

\section{A Perelman-type Entropy}

\noindent Following Perelman, we will establish an entropy by examining the spectrum of
a certain Schr\"odinger operator. Since we work with noncompact manifolds, a certain
amount of mathematical care is important. Thus certain function spaces must make an
appearance. The first is $H^1(M)$, the space of functions that are square-integrable
with respect to the metric volume element on $M$ and have square-integrable
(distributional) derivative. For $u\in H^1(M)$, $R$ falling off (or merely bounded),
and $\kappa\in \mathbb{R}$, we can define the functional
\begin{equation}
F^{(\kappa)}[g,u]:=\int_M \left ( 4\vert \nabla u\vert^2+\kappa R u^2 \right )\, dV(g)\
.
\label{eq8}\\
\end{equation}
Next we consider the subset ${\mathcal C}\subseteq H^1(M)$ of normalized ($\int u^2
dV(g)=1$) non-negative functions in $H^1(M)$. We use these to define the entropy, which
is
\begin{equation}
\lambda^{(\kappa)}(\tau) := \inf_{u\in \mathcal{C}}F^{(\kappa)}(g(\tau),u)\ .
\label{eq9}
\end{equation}
We take $\kappa\ge 1$: this, we will see, will ensure that the entropy is monotonic.
Now integrate (\ref{eq8}) by parts and impose fall-off conditions on $u$ to neglect
boundary terms. This shows that $\lambda^{(\kappa)} (\tau)$ is the left endpoint of the
spectrum of the Schr\"{o}dinger operator $-4 \Delta + \kappa R$. Our arguments will
require a discrete spectrum. Perelman worked with compact manifolds where this is
always the case, so he set $\kappa=1$. In our noncompact case, if $R<0$ somewhere,
there will be a discrete spectrum of eigenfunctions if we choose $\kappa$ large enough.
Then $\lambda^{(\kappa)}(\tau)$ will belong to the minimum eigenfunction $\bar u\in
\mathcal C$.
(For $R \geq 0$ everywhere, the spectrum is continuous and starts from zero, so
$\lambda^{(\kappa)} (\tau) = 0\ \forall\tau$: this is not a useful entropy.)

We will sometimes ``approximate'' the elements of $\mathcal C$ by
functions that are exactly $\propto 1/r^m$  near infinity,
$m>D/2$. These functions belong to ${\mathcal D}=\{ u\in {\mathcal
C} \vert u = w+k/r^m, w\in C^\infty_0, k={\rm const} \}$.
$\mathcal D$ is dense in $\mathcal C$, so continuity of the map
$H^1(M) \ni u \mapsto F^{(\kappa)} (g(t),u)\in \mathbb{R}$ gives
\begin{equation} \lambda^{(\kappa)}(\tau) = \inf_{u\in
\mathcal{D}}F^{(\kappa)} (g(\tau),u)\ , \label{eq10}
\end{equation}
which is more useful than (\ref{eq9}) for calculations.

Now say that $g(\tau)$ solves (\ref{eq3}) on $\tau\in[0,\tau_*]$.
To find $u(\tau)$, use $g(\tau)$ to write the backwards evolution
equation
\begin{equation}
\frac{\partial v}{\partial \tau} = \frac{\alpha'}{2}\left (-\Delta
v +Rv\right ) \ . \label{eq11}
\end{equation}
Solve this for some given ``initial'' data $v(\tau_*)$, where $\sqrt{v(\tau_{*})}\in
\mathcal{D}$. Such a solution always exists for $\tau\le\tau_*$; moreover, $v(\tau)>0$,
so we can define $u(\tau) := \sqrt{v(\tau)} > 0$ and $P(\tau) := -\ln(v(\tau))$. It can
be shown that
\begin{equation}
u(\tau) \equiv \sqrt{v(\tau)}\equiv e^{-P/2}\in H^{1}(M)\ , \quad
\tau \leq \tau_{*}\ , \label{eq12}
\end{equation}
so $F^{(\kappa)}[g(\tau),u(\tau)]$ is defined, and $e^{-P(\tau)}$
and the derivatives of $P$ fall off uniformly, fast enough to
allow us to discard boundary terms when integrating by parts
\cite{FallOff}. We note that $\frac{d}{d\tau} (u^{2} dV(g))
\equiv\frac{d}{d\tau} (v dV(g)) = 0$ (since from (\ref{eq3}) we
have $\frac{1}{\sqrt{g}}\frac{\partial\sqrt{g}}{\partial
\tau}=-\frac{\alpha'}{2}R$)
and $u(\tau_{*})\in \mathcal{D}$ 
implies that
\begin{equation}
\|u(\tau)\| = 1\ , \quad \tau \leq \tau_{*}\ ,  \label{eq13}
\end{equation}
so in fact $u(\tau)\in {\mathcal C}$, $\tau\le \tau_*$.
From (\ref{eq11}), $P(\tau)$ satisfies
\begin{equation}
\frac{\partial P}{\partial \tau} = \frac{\alpha^{'}}{2}(-\Delta P
+ \vert \nabla P \vert^2 - R)\ . \label{eq14}
\end{equation}

Now we are in a position to differentiate
\begin{eqnarray}
F^{(\kappa)} (\tau) &:=& F^{(\kappa)} \left ( g(\tau),e^{-P(\tau)/2}\right )\nonumber\\
&=& \int \left(|\nabla P|^2+\kappa R\right)e^{-P}dV(g)\ .
\label{eq15}
\end{eqnarray}
Integrating by parts to simplify the result, we get
\begin{eqnarray}
\frac{dF^{(\kappa)}}{d\tau}&=&\int_M \bigg \{ \left ( \nabla_i P
\nabla_j P\right ) \frac{\partial g^{ij}}{\partial \tau}+2 \big (
\vert \nabla P\vert^2\nonumber\\
&&\quad -\Delta P\big )\frac{\partial P}{\partial \tau}+\kappa
\frac{\partial R}{\partial \tau}
+\big ( \vert \nabla P \vert^2 \nonumber\\
&&\quad +\kappa R \big ) \frac{\partial}{\partial \tau}\log \left
( e^{-P}\sqrt{g} \right ) \bigg \} e^{-P} dV(g)\ .\label{eq16}
\end{eqnarray}
We could now insert the flow equations
(\ref{eq3},\ref{eq7},\ref{eq14}) into (\ref{eq16}), but the result
would not appear manifestly non-negative. To make it manifest,
recall that the flow equations are not form-invariant with respect
to $\tau$-dependent diffeomorphisms. Under the diffeomorphism
generated by $-\frac{\alpha'}{2}\nabla P$, equations (\ref{eq3}),
(\ref{eq7}), and (\ref{eq14}) become
\begin{eqnarray}
\frac{\partial}{\partial \tau} g_{ij} &=& -\alpha' \left (
R_{ij}+\nabla_i\nabla_j P\right )\label{eq17}\ ,\\
\frac{\partial R}{\partial \tau} &=& \frac{\alpha'}{2} \left (
\Delta R +2R_{ij}R^{ij} -\nabla_i R\nabla^i P \right )
\label{eq18}\ ,
\end{eqnarray}
\begin{equation}
 \frac{\partial P}{\partial \tau} =
-\frac{\alpha^{'}}{2}(\Delta P + R)\ . \label{eq19}
\end{equation}
We call these the {\em Perelman gauge} flow equations. But
integrals over $M$, such as $F^{(\kappa)}$, are invariant under
diffeomorphisms. If the diffeomorphism preserves the asymptotic
structure, we remain justified in discarding boundary terms, so
(\ref{eq16}) holds in such a gauge. Inserting the Perelman gauge
flow equations into (\ref{eq16}), integrating by parts, and using
the boundary conditions, the Ricci identity, and the contracted
second Bianchi identity, we get
\begin{eqnarray}
\frac{dF^{(\kappa)}}{d\tau}&=& \alpha^{'} \int_M \left(\left \vert R_{ij}
+\nabla_i\nabla_j P\right \vert^2+\left ( \kappa -1\right )
\left\vert R_{ij}\right\vert^2 \right)\nonumber\\
&&\qquad e^{-P} dV(g)\ , \label{eq20}
\end{eqnarray}
which is manifestly non-negative if $\kappa\ge 1$ (which we take
from here on). Hence $\gamma(\tau) \leq \gamma(\tau_{*})$ for
$\tau \leq \tau_{*}$. In other words we have
\begin{equation}
F^{(\kappa)}(g(\tau),u(\tau)) \leq F^{(\kappa)}(g(\tau_{*}),u(\tau_{*})) \label{eq21}
\end{equation}
for $\tau \leq \tau_{*}$. Using $u(\tau)> 0$, (\ref{eq12}), and (\ref{eq13}) we see
that $u(\tau)\in \mathcal{C}$ and hence by the definition of the entropy and
(\ref{eq21}) it follows that $\lambda^{(\kappa)}(\tau) \leq F^{(\kappa)}
(g(\tau_{*}),u(\tau_{*}))$ for $\tau \leq \tau_{*}$. But recall that $u(\tau_{*})$ was
taken to be an arbitrary element of $\mathcal{D}$ and hence we get by (\ref{eq10}) that
\begin{equation}
\lambda^{(\kappa)}(\tau) \leq  \lambda^{(\kappa)}(\tau_{*}) \quad {\rm\ for\ }\tau \leq
\tau_{*} \label{eq22}
\end{equation}
which proves that the \emph{entropy is increasing}.

\section{No Breathers I: $R<0$ somewhere}

\noindent If $g$ is a breather, then $\lambda^{(\kappa)}(\tau_1) =
\lambda^{(\kappa)}(\tau_2)=:\Lambda$. Because entropy is monotonic, then
$\lambda^{(\kappa)}(\tau)=\Lambda$ for all $\tau\in[\tau_1,\tau_2]$. We must now show
that this statement has geometrical consequences. The trick is to construct a function
$u={\bar u}(\tau)$ that realizes the entropy $\Lambda$ at all $\tau$. Now if $R<0$
somewhere, we can choose $\kappa$ large enough so that there is a minimizer
$\bar{u}(\tau_2)\in \mathcal{C}$ for the entropy realizing
$\lambda^{(\kappa)}(\tau_{2})$:
\begin{equation}
\lambda^{(\kappa)}(\tau_2) = F^{(\kappa)}(g(\tau_2),\bar{u}(\tau_{2}))\ . \label{eq23}
\end{equation}
We choose a sequence $u_{n}(\tau_{2})\in \mathcal{D}$ that converges to
$\bar{u}(\tau_2)$ in $H^{1}(M)$ and, as above, use the squares
$v_n(\tau_2):=u_n^2(\tau_2)$ as ``initial'' data at $\tau_2$ for a sequence of
solutions $v_{n}(\tau):=v(\tau)$ of (\ref{eq11}) on $\tau\in[\tau_1,\tau_{2}]$ (where
now $\tau_{*}=\tau_{2}$). Because we start with $u_{n}(\tau_{2})\in \mathcal{D}$, we
are assured that $u_{n}(\tau) := \sqrt{v_{n}(\tau)}$ is defined (and in $\mathcal C$),
and so
\begin{equation}
F^{(\kappa)}_n(\tau) := F^{(\kappa)}(g(\tau),u_{n}(\tau))\ ,\quad
\tau\in[\tau_1,\tau_{2}] \label{eq24}
\end{equation}
is also defined. Repeating the calculations leading to (\ref{eq20}) (using
$P_{n}(\tau):= -\ln(v_{n}(\tau))$), we obtain
\begin{equation}
\frac{dF^{(\kappa)}_n}{d\tau} \ge 0 \, , \label{eq25}
\end{equation}
and therefore $F^{(\kappa)}_n(\tau) \leq F^{(\kappa)}_n(\tau_2)$.
Putting everything together, we have for $\tau \in
[\tau_1,\tau_2]$ that the breather obeys
\begin{eqnarray}
\Lambda&=&\lambda^{(\kappa)}(\tau_{2})=\lambda^{(\kappa)}(\tau_1)
\leq \lambda^{(\kappa)}(\tau)\nonumber\\
&\leq& F^{(\kappa)}_n(\tau) \leq
F^{(\kappa)}_{n}(\tau_2)\to\lambda^{(\kappa)}(\tau_2) \label{eq26}
\end{eqnarray}
as $n\to\infty$. In particular, this shows that
$F^{(\kappa)}_n(\tau)\to\lambda^{(\kappa)}(\tau)=\Lambda$ for each $\tau \in
[\tau_1,\tau_2]$, so $u_{n}(\tau)$ is a \emph{minimizing sequence}. Thus, passing to a
subsequence if necessary, we have $u_{n}(\tau) \rightarrow \bar{u}(\tau)$ weakly in
$H^1(M)$ with $\bar{u}(\tau) \in \mathcal{C}\subseteq H^{1}(M)$ and
$\lambda^{(\kappa)}(\tau) = F^{(\kappa)}(g(\tau),\bar{u}(\tau))$, as desired.

Next, integrating (\ref{eq25}) yields
$\int_{\tau_1}^{\tau_2}\frac{dF^{(\kappa)}_n}{d\tau}\, d\tau =
F^{(\kappa)}_{n}(\tau_2)-F^{(\kappa)}_{n}(\tau_1) \rightarrow 0$ as $n\rightarrow
\infty$. Since $\frac{dF^{(\kappa)}_n}{d\tau}\ge 0$, then it follows that, once again
passing to a subsequence if necessary, $\lim_{n\rightarrow \infty}
\frac{dF^{(\kappa)}_n}{d\tau}= 0$ pointwise on $(\tau_{1},\tau_{2})$ except perhaps on
a set $S$ of measure zero. Then (\ref{eq20}) (with $\kappa$ large and $e^{-P}=u_n^2$)
implies that $\int_{M}\vert R_{ij}\vert^{2}u_{n}^{2}dV \rightarrow 0$ as $n\rightarrow
\infty$ for each $\tau\notin S$. It follows that $\vert R_{ij}\vert u_{n}\rightarrow 0$
in $H^{1}(M)$, $\tau\notin S$. But we also know that $\vert R_{ij}\vert u_{n}
\rightarrow \vert R_{ij}\vert \bar{u}$ weakly in $H^{1}(M)$. By the uniqueness of weak
limits and the fact that $\bar{u}>0$ we get that $R_{ij}(\tau)=0$ for each $\tau\in
\Sigma$. By continuity in $\tau$, $R_{ij}(\tau)=0$ for all $\tau \in
[\tau_{1},\tau_{2}]$. Since we assume $R<0$ somewhere, this is a contradiction.

\section{No Breathers II: $R\ge 0$}

\noindent We now assume that $0\le R(\tau)$ and that $\int_M R(\tau)<\infty$ for all
$\tau \in [\tau_{1},\tau_2]$. A wide class of local solutions satisfy this condition:
in fact, this is what is needed to ensure that the mass of the manifold is well-defined
and unique. However, for this case the entropy defined above is always zero and does
not rule out breathers. We therefore need a new entropy.

Since the scalar curvature satisfies (\ref{eq7}), for which the
minimium principle applies, it follows that $R(\tau_{1})\geq 0$
implies $R(\tau) \geq 0$ all $\tau \geq \tau_{1}$. Thus if the
scalar curvature is initially non-negative then it stays
non-negative along the flow. The volume element obeys
$\frac{\partial}{\partial \tau} dV(g) = -\frac{\alpha'}{2}RdV(g)$.
Integrating this equation yields
\begin{equation}
\sqrt{g(\tau)}-\sqrt{g(\tau_1)} =
-\frac{\alpha'}{2}\int_{\tau_{1}}^{\tau} R(s)\sqrt{g(s)}ds \ .
\label{eq27}
\end{equation}
Since $\int_MR(\tau)\sqrt{g(\tau)} d^Dx$ is bounded on
$\tau\in[\tau_1,\tau_2]$, it follows that $\int_{\tau_{1}}^{\tau}
R(s)\sqrt{g(s)}ds$ is integrable over $M$ and hence the new
entropy
\begin{eqnarray}
\mu(\tau) &:=& -\int_{M}\left ( \sqrt{g(\tau)}-\sqrt{g(\tau_1)}\right )
d^Dx \nonumber\\
&=& \frac{\alpha'}{2}\int_{M}\int_{\tau_{1}}^{\tau} R(s)\sqrt{g(s)}ds d^Dx \nonumber\\
&=& \frac{\alpha'}{2}\int_{\tau_{1}}^{\tau} \int_{M} R(s)dV(g(s))ds \label{eq28}
\end{eqnarray}
is finite where, for simplicity, we have taken $M=\mathbb{R}^D$.
Now, $R(\tau)\geq 0$ implies that $\mu(\tau)$ is nondecreasing.
Moreover we have that $\mu(\tau_{2}) > \mu(\tau_{1}) = 0$ unless
$R(\tau) = 0$ for all $\tau \in [\tau_{1},\tau_{2}]$. If $R(\tau)
= 0$ for all $\tau \in [\tau_{1},\tau_{2}]$, then from (\ref{eq7})
we see that $R_{ij}(\tau) = 0$ for all $\tau$ which shows that
$g(\tau_{1})$ must be a fixed point. So now let us assume that
there exists a $\tau \in [\tau_{1},\tau_{2}]$ for which $R(\tau)
\neq 0$. Then
\begin{equation}
\mu(\tau_{2}) > \mu(\tau_{1}) = 0 \ . \label{eq29}
\end{equation}
Let $G$ denote $\det(g_{ij}(\tau_{2}))$. Then, applying the
breather condition (\ref{eq6}) to (\ref{eq28}), we get
\begin{equation}
\mu(\tau_{2}) = -\frac{\alpha'}{2}\int_{M} \left [\sqrt{G}-
\det(J(\phi))\sqrt{G}\circ\phi \right ] d^Dx \ \label{eq30},
\end{equation}
where $J(\phi)$ is the Jacobian matrix, i.e., $J(\phi)^{i}{}_{j} :=
\partial_{j}\phi^{i}$. To proceed, we assume that the
diffeomorphism $\phi$ lies in the connected component of the identity so that there
exists a 1-parameter family of diffeomorphisms $\psi_{t}$ $(0\leq t \leq 1)$ such that
$\psi_{0}=\rm{id}$, $\psi_{1} = \phi$, and $\psi_{t}$ preserves the asymptotic
structure for each $t\in[0,1]$. Now consider the Lagrangian density
\begin{equation}
\mathcal{L}(\psi^{i}_{t},\partial_{j}\psi^{i}) :=
\det(J(\psi_{t}))\sqrt{G}\circ\psi_{t} - \sqrt{G}\ .\label{eq31}
\end{equation}
A straightforward calculation shows that the Euler-Lagrange equations are automatically
satisfied; that is, we have the identity
\begin{equation}
-\partial_{j}\frac{\partial
\mathcal{L}}{\partial\partial_{j}\psi_{t}^{i}} + \frac{\partial
\mathcal{L}}{\partial \psi_{t}^{i}} = 0 \ . \label{eq32}
\end{equation}
Thus if we define
\begin{equation}
I(t) := \frac{\alpha'}{2}\int_{M} \mathcal{L}(\psi_{t}^{i},\partial_{j}\psi_{t}^{i})
d^Dx\ , \label{eq33}
\end{equation}
then differentiating $I(t)$, integrating by parts, and using
(\ref{eq32}) yields
\begin{equation}
\frac{dI}{dt} = \frac{\alpha'}{2}\int_{S_{\infty}}\det(J(\psi_{t}))J(\psi_{t})^{-1\,
j}{}_{i} \frac{d\psi^{i}_{t}}{dt} \,   dS_{j}\ . \label{eq34}
\end{equation}
If $\psi_{t}$ approaches the identity as $r \rightarrow \infty$ and
$\int_{S_{\infty}}d\psi^{i}_{t}/dt\, dS_{i} = 0$ then we get that $I(t)=\rm{const}$.
But $I(1)=\mu(\tau_{2})$, and $I(0)= 0$, so $\mu(\tau_{2})=0$. This shows that if the
diffeomorphism $\phi$ has the form $\phi^{i}(x) = x^{i}+\bar{\phi}^{i}(x)$ where
$\int_{S_{\infty}}\bar{\phi^{i}}\,dS_{i}=0$ then we must have $\mu(\tau_{2})=0$ which
contradicts (\ref{eq29}). Therefore if $g(\tau_1) = \phi^{*}g(\tau_2)$, then the
diffeomorphism will, at least, violate
\begin{equation}
\int_{S_{\infty}} (\phi^{i}(x) - x^{i})\,dS_{i} = 0 \ .
\label{eq35}
\end{equation}
Whether there exist solutions periodic modulo diffeomorphisms that
violate this condition is an interesting question which deserves
attention.

In closing, we ask, can we pass now to RG flows wherein the second-order corrections
are important? Unfortunately, in such circumstances, we do not know whether the flow
will always preserve asymptotic flatness, even for short times. If it does, then it
still may not have other necessary properties, such as the preservation of positive
scalar curvature upon which the volume entropy argument depends. This is illustrative
of the potential difficulties in generalizing our arguments to the second-order case.

\section{Acknowledgments}

\noindent This work was partially supported by a grant from the Natural Sciences and
Engineering Research Council of Canada. VS was supported by a postdoctoral fellowship
from the Pacific Institute for the Mathematical Sciences. We thank Joe Polchinski for
discussions and comments and Bruce Campbell for bringing reference \cite{Friedan} to
our attention.

\section{Appendix}

\noindent Here we give distinct geometric arguments for the case where $R < 0$
somewhere. Our arguments closely parallel those of Ivey \cite{Ivey} for compact
manifolds. By asymptotic flatness, $R\to 0$ uniformly in $\tau$ at spatial infinity.
Since the flow is smooth, $R$ has an infimum, say $-k$, by assumption negative. Outside
a big enough compact set $C:=[0,T]\times K$ asymptotic flatness guarantees that
$R>-k/2$. Thus the infimum must be approached inside $C$, which is compact, so $R$
achieves the minimum value $-k$ (in $C$ and thus in $[0,T]\times M$). Moreover, if the
solution is a breather with period $<T$, the minimum must be achieved at least once in
the open region $(0,T)\times M$. But if $R$ has a minimum in $(0,T)\times M$, then
$\frac{\partial R}{\partial t}=0$ and $\Delta R\ge 0$ there, so we conclude from
(\ref{eq7}) (or (\ref{eq18})---the argument is gauge invariant) that $R_{ij}$ vanishes
there. Taking the trace, $R=0$ at the minimum. This is a contradiction, so there cannot
be a breather with period $<T$ and somewhere negative scalar curvature. But $T$ is
arbitrary.

To conclude, consider the case of $R\ge 0$ with $R=0$ at an isolated point. The
argument given by Ivey \cite{Ivey} for this case is essentially local and carries over
to the case of asymptotic flatness (though we constructed our own version to verify
details). The basic idea is that (\ref{eq7}) obeys a Hopf lemma, which gives that
$dR\neq 0$ if $R=0$ at an isolated point with $\tau>0$. Since $dR$ must be zero at an
interior minimum, either we never have $R=0$ for $\tau>0$, or $R=0$ everywhere and
throughout the flow. In the latter case, we see from (\ref{eq7}) that $R_{ij}=0$
everywhere and the breather is a fixed point of the flow.



\begin{thebibliography}{100}

\bibitem{Z} AB Zamolodchikov, Pis'ma Zh Eksp Teor Fiz 46 (1987) 129
[Sov Phys JETP Lett 46 (1987) 129].
\bibitem{Pol} J Polchinski, Nucl Phys B303 (1988) 226.
\bibitem{niemi} A Morozov and AJ Niemi, Nucl Phys B666 (2003) 311 [hep-th/0304178].
\bibitem{APS} A Adams, J Polchinski, and E Silverstein,
JHEP 0110 (2001) 029 [hep-th/0108075].
\bibitem{GHMS} M Gutperle, M Headrick, S Minwalla, and V Schomerus,
JHEP 0301 (2003) 073 [hep-th/0211063].
\bibitem{HMT} M Headrick, S Minwalla and T Takayanagi, preprint 2004 [hep-th/0405064]
and references therein.
\bibitem{Bakas} I Bakas, Fortsch Phys 52 (2004) 464.
\bibitem{Tseytlin} AA Tseytlin, Phys Lett B194 (1987) 63.
\bibitem{EH} G Ecker and J Honerkamp, Nucl Phys B35 (1971) 481.
\bibitem{Friedan} DH Friedan, Ph D thesis, University of California,
Berkeley, 1980 (unpublished); Phys Rev Lett 45 (1980) 1057.
\bibitem{FT} ES Fradkin and AA Tseytlin, Phys Lett B158 (1985)
316; CG Callan, D Friedan, EJ Martinec and MJ Perry,
Nucl Phys B262 (1985) 593; and see also {\em String theory} by
J. Polchinski, Cambridge University Press for further references.
\bibitem{Hamilton} RS Hamilton, J Diff Geom 17 (1982) 255.
\bibitem{DEbook} J Jost, {\it Partial Differential Equations} (Springer, New York, 2002), p 77ff.
\bibitem{Perelman} G Perelman, preprint 2002 [math.DG/0211159].
\bibitem{Ivey} T Ivey, Diff Geom Appl 3 (1993) 301.
\bibitem{FallOff} The fall-off rates are $e^{-P(\tau)} \in \mathcal{O}(1/r^{2m-\epsilon})$,
$\partial^{I}P(\tau) \in \mathcal{O} (1/r^{|I|-\epsilon})$,
$\partial^{I}\frac{\partial}{\partial \tau} P(\tau) \in \mathcal{O}
(1/r^{|I|+2-\epsilon})$, uniformly in $\tau$ for any $\epsilon > 0$, and $I$ a
multi-index with $1\leq |I|\leq 2$.
\end{thebibliography}
\end{document}